\documentclass[aps,prl,twocolumn,nofootinbib]{revtex4-1}

\usepackage[usenames,dvipsnames]{xcolor}
\usepackage{hyperref}
\hypersetup{
	colorlinks=true,
	urlcolor=Maroon,
	linkcolor=RoyalBlue,
	citecolor=Maroon,
	bookmarks=true,
	pdftitle={Kinematic Jacobi Identity is a Residue Theorem: Geometry of Color-Kinematics Duality for Gauge and Gravity Amplitudes},
	pdfauthor={Sebastian Mizera},
	pdfdisplaydoctitle=true,
	pdfstartview=FitH
}

\usepackage{amsmath}
\usepackage{amsfonts}
\usepackage{mathtools}
\usepackage{microtype}
\usepackage{bm}
\usepackage[utf8]{inputenc}
\usepackage{blkarray}
\usepackage{bigstrut}
\usepackage{nccmath}
\usepackage{enumitem}
\usepackage{braket}
\usepackage{dsfont}
\usepackage[export]{adjustbox}
\usepackage[english]{babel}

\allowdisplaybreaks
\graphicspath{{figures/}}

\DeclareMathOperator{\Res}{Res}
\DeclareMathOperator{\Pf}{Pf}
\DeclareMathOperator{\SL}{SL}

\DeclareMathOperator{\Tr}{Tr}

\def \be {\begin{equation}}
\def \ee {\end{equation}}
\def \nn {\nonumber}
\def \la {\left<}
\def \ra {\right>}

\def \C  {\mathbb{C}}
\def \CP  {\mathbb{CP}}
\def \Z  {\mathbb{Z}}

\begin{document}

\title{Kinematic Jacobi Identity is a Residue Theorem:\\ Geometry of Color-Kinematics Duality for Gauge and Gravity Amplitudes}
\author{Sebastian Mizera}
\email{smizera@ias.edu}
\affiliation{Institute for Advanced Study, Einstein Drive, Princeton, NJ 08540, USA}

\begin{abstract}\noindent
We give a geometric interpretation of color-kinematics duality between tree-level scattering amplitudes of gauge and gravity theories. Using their representation as intersection numbers we show how to obtain Bern--Carrasco--Johansson numerators in a constructive way as residues around boundaries of the moduli space. In this language the kinematic Jacobi identity between each triple of numerators is a residue theorem in disguise.
\end{abstract}

\maketitle

\section{Introduction}

Computation of scattering amplitudes in gravitational theories has traditionally posed a formidable task---even for tree-level processes---due to a proliferation of Feynman diagrams involved. This fact has changed with the introduction of \emph{color-kinematics duality} \cite{Bern:2008qj,Bern:2010ue} by Bern, Carrasco, and Johansson (BCJ), which provides a shortcut in computing gravitational observables by extracting the relevant information from gauge theory. It has since found applications in a spectrum of topics ranging from the study of ultraviolet properties of gravity \cite{Bern:2011rj,Bern:2012uf,Bern:2012cd,Bern:2012gh,Bern:2014lha,Bern:2018jmv}, through the construction of classical solutions \cite{Monteiro:2014cda,Ridgway:2015fdl,Goldberger:2016iau,Luna:2016hge,Bahjat-Abbas:2017htu,Plefka:2018dpa,Adamo:2018mpq,Luna:2018dpt}, to gravitational-wave physics \cite{Cheung:2018wkq,Kosower:2018adc,Bern:2019nnu,Antonelli:2019ytb,Bautista:2019tdr,Bern:2019crd}.

Working at tree level, let us make the statement of color-kinematics duality more precise. Scattering amplitudes of $n$ gauge bosons can be expressed as
\be\label{intro-gauge}
{\cal A}_n^{\text{gauge}} = \sum_{\Gamma} \frac{n_\Gamma\, c_\Gamma}{\prod_{e\in \Gamma} p_e^2},
\ee
where the sum goes over all $(2n{-}5)!!$ trivalent trees $\Gamma$ with propagators $p_e^2$ associated to each internal edge $e$ of $\Gamma$. Here $c_\Gamma$ denotes the color structure attached to each diagram, while $n_\Gamma$ is the remaining part of the numerator involving kinematic information such as contractions of momenta and polarization vectors.

Let us isolate triples of terms in \eqref{intro-gauge} with graphs denoted by $(\Gamma_{\!s},\Gamma_{\!t},\Gamma_{\!u})$ differing only by a single subdiagram as follows:
\be\label{triplet}
\includegraphics[scale=1,valign=c]{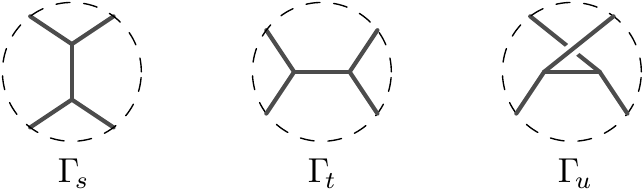}
\ee
Color structures associated to such triples satisfy the Lie algebra Jacobi identity, $c_{\Gamma_{\!s}} + c_{\Gamma_{\!t}} + c_{\Gamma_{\!u}} = 0$.
Suppose that for every $(\Gamma_{\!s},\Gamma_{\!t},\Gamma_{\!u})$ we enforce a similar condition on the kinematic numerators,
\be\label{KJR}
n_{\Gamma_{\!s}} + n_{\Gamma_{\!t}} + n_{\Gamma_{\!u}} =0,
\ee
known as the \emph{kinematic Jacobi identity}. Since the numerators coming from Feynman diagram expansion do not naturally satisfy \eqref{KJR}, it is typically a difficult task to bring them into such a form by reshuffling terms in \eqref{intro-gauge}. Assuming this can be done BCJ proposed \cite{Bern:2008qj} that scattering amplitudes in gravity theory can be written, up to normalization, as
\be\label{intro-gravity}
{\cal A}_n^{\text{gravity}} = \sum_{\Gamma} \frac{n_\Gamma\, \tilde{n}_\Gamma}{\prod_{e\in \Gamma} p_e^2},
\ee
where $n_\Gamma$'s and $\tilde{n}_\Gamma$'s are two (possibly distinct) sets of Jacobi-satisfying numerators. This statement is now proven \cite{Bern:2010yg} and can be extended to loop level \cite{Bern:2008qj,Bern:2010ue,Bern:2010tq,Du:2012mt,Boels:2013bi,Bjerrum-Bohr:2013iza,Bern:2013yya,Ochirov:2013xba,Mafra:2015mja,Mogull:2015adi,He:2015wgf,He:2017spx,Geyer:2017ela,Geyer:2019hnn}, gauge and gravity theories with different supersymmetry and matter content \cite{Broedel:2012rc,Carrasco:2012ca,Johansson:2014zca,Chiodaroli:2014xia,Chiodaroli:2015rdg,Chiodaroli:2015wal,Johansson:2017srf,Chiodaroli:2017ehv,Chiodaroli:2018dbu}, as well as various other theories \cite{Cachazo:2014xea,Cachazo:2016njl,Du:2016tbc,Carrasco:2016ldy,Cheung:2016prv,Carrasco:2016ygv,Elvang:2018dco,Mizera:2018jbh}. Kinematic algebras leading to \eqref{KJR} have been investigated in \cite{Monteiro:2011pc,BjerrumBohr:2012mg,Chen:2019ywi}. For a comprehensive review of color-kinematics duality see \cite{Bern:2019prr}.

At this stage one can ask if the kinematic Jacobi identity \eqref{KJR} has a geometric interpretation, and whether there exists a representation of scattering amplitudes that manifests this fact. These questions turn out to have a common answer, whose elucidation is the goal of this letter.

It has recently emerged that a natural framework for addressing such problems is that of intersection theory \cite{Mizera:2017rqa}. It was previously used to provide a geometric interpretation of Kawai--Lewellen--Tye (KLT) \cite{Kawai:1985xq} relations between string- and field-theory amplitudes in terms of intersections of associahedra \cite{Mizera:2017cqs,Mizera:2016jhj,Mizera:2019gea}; write down higher-loop monodromy and BCJ \cite{Plahte:1970wy,Bern:2008qj} relations for loop integrands \cite{Casali:2019ihm}; understand precise conditions under which the low-energy limit of string-theory amplitudes localizes on scattering equations \cite{Mizera:2017rqa,Mizera:2019gea}; as well as give a new perspective on differential equations, dimensional recurrence relations, and integration-by-parts identities for multi-loop Feynman integrals \cite{Mastrolia:2018uzb,Frellesvig:2019kgj,Frellesvig:2019uqt,Mizera:2019vvs}, among other applications \cite{delaCruz:2017zqr,Frost:2018djd,Li:2018mnq,Brown:2019jng,Brown:2019wna,Abreu:2019wzk,Kalyanapuram:2019nnf}. At the same time this line of research unraveled connections between scattering amplitudes and more formal topics including Morse theory \cite{Mizera:2017rqa}, Euler characteristics \cite{Mastrolia:2018uzb,Mizera:2019gea}, Landau--Ginzburg models \cite{Mizera:2019vvs}, and Yang--Baxter relations \cite{Mizera:2019gea}.

The central role in this theory is played by the so-called \emph{intersection numbers}, which provide a geometric representation of tree-level amplitudes in various quantum field theories \cite{Mizera:2017rqa,Mizera:2019gea}. Selecting a theory amounts to specifying two differential forms, $\varphi_-$ and $\varphi_+$, on the moduli space of Riemann spheres with $n$ punctures, ${\cal M}_{0,n}$. In the low-energy limit intersection numbers are computed by \cite{Mizera:2019gea}
\be\label{intersection-number}
\sum_{\Gamma} \frac{\Res_{v_\Gamma}(\varphi_-) \Res_{v_\Gamma}(\varphi_+)}{\prod_{e\in \Gamma} p_e^2}.
\ee
Here the sum is of exactly the same form as in \eqref{intro-gauge} and \eqref{intro-gravity}, and the role of numerators---both color and kinematic ones---is played by the residues around maximal-codimension boundaries of the moduli space, $v_\Gamma$, which are in one-to-one map with trivalent diagrams $\Gamma$.

We will prove that the numerators in \eqref{intersection-number} always satisfy the kinematic Jacobi identity \eqref{KJR} as a consequence of a residue theorem, thus providing a manifestly color-kinematics dual representation of amplitudes.

In this language the problem of finding numerators for various theories translates to different choices of $\varphi_{\pm}$. After reviewing a known catalog of such forms for gauge and gravity theories we give explicit examples of computing Jacobi-satisfying numerators.

\section{Boundaries and Residues}

Let us briefly review the factorization structure of the moduli space ${\cal M}_{0,n}$ provided by its compactification \cite{Deligne1969}. When a subset $R$ of punctures collides on the Riemann sphere, the surface should be thought of as ``bubbling'' into two new spheres, where an emergent puncture $I$ separates the set $R$ from the complementary set $L$ (with sizes $2 \leq |L|,|R| \leq n{-}2$): 
\be
\includegraphics[scale=1,valign=c]{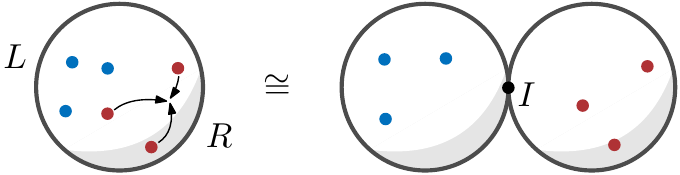}
\ee
It is a codimension-one component of the boundary divisor $\partial {\cal M}_{0,n}$. We can make this procedure concrete on the level of differential forms. Take $\varphi$ to be a top (degree $n{-}3$) holomorphic form on ${\cal M}_{0,n}$, i.e., proportional to the $\SL(2,\C)$-covariant measure
\be
d\mu_n = (z_p{-}z_q) (z_q{-}z_r) (z_p{-}z_r) \!\!\!\bigwedge_{\substack{i=1\\ i \neq p,q,r}}^{n}\!\!\! dz_i,
\ee
where $(z_p,z_q,z_r)$ denote the positions of three arbitrary punctures fixed by the action of $\SL(2,\C)$. For massless scattering we must require that $\varphi$ is invariant under $\SL(2,\C)$ transformations $z_i \mapsto (\text{A} z_i {+} \text{B})/(\text{C} z_i {+} \text{D})$ with $\text{AD}{-}\text{BC}=1$ for all $z_i$'s.

A standard way of modeling the above factorization is to embed the original sphere $\CP^1$ as a conic in $\CP^2$ with a new parameter $\epsilon$, such that it factors into $\CP^1 {\times} \CP^1$ as $\epsilon \to 0$, see, e.g., \cite{Cachazo:2012pz}. In coordinates, we perform the change of variables
\be
z_i = \begin{dcases}
	\epsilon / x_i \quad &\text{for } i \in L, \\
	y_i / \epsilon \quad &\text{for } i \in R,
	\end{dcases}
\ee
where $x_i$'s and $y_i$'s are positions of punctures on the new spheres with exactly two $x_i$'s and two $y_i$'s fixed. Since the boundary lies along $\{\epsilon^2{=}0\}$ we can simply take
\be\label{residue}
\Res_{\epsilon^2 = 0} (\varphi) = \varphi_L \wedge \varphi_R,
\ee
where $\varphi_L(x_i)$ and $\varphi_R(y_i)$ are now top (degree $|L|{-}2$ and $|R|{-}2$) holomorphic forms on the moduli spaces ${\cal M}_{0,|L|+1}$ and ${\cal M}_{0,|R|+1}$ of the left and right sphere respectively. From the perspective of the particles on the left sphere the emergent puncture is at $x_I = 0$, while from the right sphere it is at $y_I = 0$. In the special case of two punctures colliding, i.e., $R=\{z_i, z_j\}$ the residue becomes simply $\Res_{z_i = z_j}(\varphi) = \varphi_L$ up to orientation. Intuitively, one might think of \eqref{residue} as extracting a singular part in the operator product expansion between operators from the set $R$ being replaced by $I$ (or those from $L$ being replaced by $I$ from the other sphere's perspective)

Repeating this procedure exactly $n{-}3$ times one obtains maximal-codimension components (vertices) $v_\Gamma$ of $\partial {\cal M}_{0,n}$, which are in one-to-one map with trivalent graphs $\Gamma$, as all punctures are fixed by the action of $\text{SL}(2,\C)$, e.g.,
\be
\includegraphics[scale=1,valign=c]{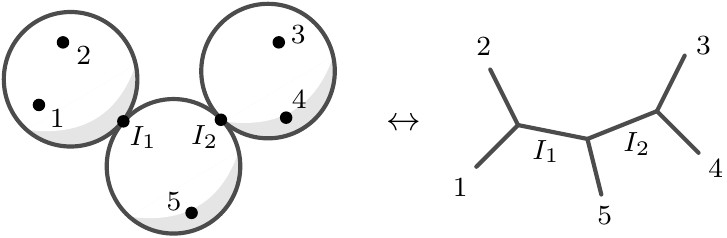}
\ee
The corresponding numerator $n_\Gamma = \Res_{v_\Gamma}(\varphi)$ is a function computed by applying \eqref{residue} consecutively $n{-}3$ times.

There exists an alternative way of computing $\Res_{v_\Gamma}(\varphi)$, based on the dihedral extension of ${\cal M}_{0,n}$ employing cross-ratio coordinates suited for each $v_\Gamma$ \cite{Brown:2009qja}, which is particularly useful for planar amplitudes, see, e.g., \cite{Arkani-Hamed:2017mur,delaCruz:2017zqr,Mizera:2019gea}.

\section{Kinematic Jacobi Identity}

Let us consider the stage at which bubbling already happened $n{-}4$ times, i.e., when we are only one residue away from a trivalent factorization. It means there is exactly one sphere with four punctures:
\be\label{middle-sphere}
\includegraphics[scale=1,valign=c]{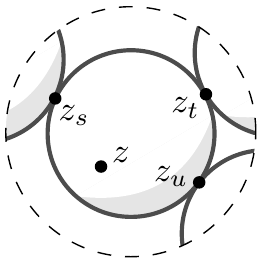}
\ee
This leaves us with a one-form $\varphi_M$ on the moduli space of the ``middle'' sphere, which was computed as an $(n{-}4)$-fold residue of the original form $\varphi$. Let us call the unfixed puncture $z$ and the fixed ones $(z_s, z_t, z_u)$, such that $z$ colliding with $z_i$ leads to a trivalent graph $\Gamma_{i}$, as in \eqref{triplet}. By definition of the numerators entering \eqref{intersection-number} we have:
\begin{gather}
n_{\Gamma_{\!s}} = \Res_{z = z_s}(\varphi_M),\quad
n_{\Gamma_{\!t}} = \Res_{z = z_t}(\varphi_M),\nn\\
n_{\Gamma_{\!u}} = \Res_{z = z_u} (\varphi_M),
\end{gather}
which are residues around the boundaries of the remaining moduli space:
\be
\includegraphics[scale=1,valign=c]{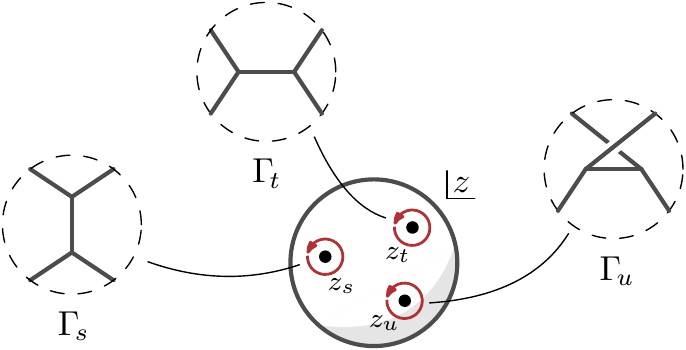}
\ee
Since there are no other poles the residue theorem reads
\be\label{KJR2}
n_{\Gamma_{\!s}} + n_{\Gamma_{\!t}} + n_{\Gamma_{\!u}} = 0,
\ee
which is precisely the kinematic Jacobi identity \eqref{KJR}. Given that we could have started with any configuration \eqref{middle-sphere}, this identity is satisfied for all possible triples $(\Gamma_{\!s}, \Gamma_{\!t}, \Gamma_{\!u})$.\footnote{The identity \eqref{KJR2} means that for each triple only two out of three numerators are $\Z$-independent. One can ask how these relations combine for subdiagrams with $m{\geq}4$ external legs by considering the ``middle'' sphere \eqref{middle-sphere} to have $m$ points. The results of \cite{arnold1969cohomology} show that all residue theorems must reduce the number of $\Z$-independent numerators down to $\dim H^{m-3}({\cal M}_{0,m},\Z) = (m{-}2)!$ from the total of $(2m{-}5)!!$.}

\section{Building Blocks}

At this stage we have demonstrated that any rational form $\varphi$ on ${\cal M}_{0,n}$ leads to Jacobi-satisfying numerators, however it does \emph{not} yet mean that the resulting \eqref{intersection-number} is a scattering amplitude. We need to learn how to pick differential forms of physical relevance, which is a domain of intersection theory.

The first step is to realize that such forms should be really treated as elements of cohomology (equivalence) classes labeled by a $\pm$ sign,
\be\label{cohomology-relation}
\varphi_\pm \;\sim\; \varphi_\pm + (d \pm dW\wedge) \xi
\ee
for any rational $(n{-}4)$-form $\xi$. Here $W$ is a potential given by
\be
W = \frac{1}{\Lambda^2} \sum_{i < j} 2 p_i {\cdot} p_j  \log (z_i {-} z_j).
\ee
with a mass scale $\Lambda$.
This is precisely how $\varphi_\pm$ ``know'' about physics through the kinematic invariants $p_i {\cdot} p_j$. To distinguish them from ordinary differential forms we call $\varphi_{\pm}$ \emph{twisted forms}. Their space is $(n{-}3)!$-dimensional \cite{aomoto1987gauss}, in contrast with the space of ordinary forms, which is $(n{-}2)!$-dimensional \cite{arnold1969cohomology}. In order to make the statements below non-trivial we typically impose that twisted forms have no kinematic poles, which in turn implies that the numerators $n_{\Gamma}$ are local.

One can construct a bilinear of $\varphi_-$ and $\varphi_+$ called their \emph{intersection number}, $\la \varphi_- | \varphi_+ \ra_{dW}$, given by integrating the two forms over the moduli space. While such invariants have been known in mathematics for decades \cite{zbMATH03996010,cho1995,matsumoto1998}, only recently they were identified as representations of tree-level scattering amplitudes in various massive and massless quantum field theories in arbitrary space-time dimension \cite{Mizera:2017rqa}, see \cite{Mizera:2019gea} for a comprehensive introduction. We focus on massless external states, $p_i^2 {=} 0$, from now on.

There exists a catalog of twisted forms, which can be mixed and matched to compute different amplitudes \cite{Mizera:2019gea}. For theories with color degrees of freedom $T^{c_i}$ we have
\be\label{varphi-color}
\varphi_{\pm}^{\text{color}} = d\mu_n \left( \frac{\Tr(T^{c_1} T^{c_2} \cdots T^{c_n})}{(z_1{-}z_2) (z_2{-}z_3) \cdots (z_n{-}z_1)} + \text{perm.} \right),
\ee
where the symmetrization involves $(n{-}1)!$ cyclic permutations (the definition is the same for both $\pm$). By construction the associated numerator is precisely the color structure of a given diagram, i.e., $\Res_{v_\Gamma}(\varphi_{\pm}^{\text{color}}) = c_\Gamma$, as in \eqref{intro-gauge}. For theories with polarization vectors $\varepsilon_i^\mu$ we can use
\be\label{varphi-gauge}
\varphi^{\text{gauge}}_{\pm} = d\mu_n\! \int \prod_{i=1}^{n} d\theta_i d\tilde{\theta}_i \frac{\theta_k \theta_\ell}{z_k {-} z_\ell} \exp \sum_{i \neq j} \Phi_{ij},\\
\ee
(the choice of $k$ and $\ell$ is arbitrary) with
\be
\Phi_{ij} = -\frac{\theta_i \theta_j p_i {\cdot} p_j + \tilde{\theta}_i \tilde{\theta}_j \varepsilon_i {\cdot} \varepsilon_j + 2(\theta_i {-} \theta_j) \tilde{\theta}_i \varepsilon_i {\cdot} p_j }{z_i {-} z_j \mp \Lambda^2 \theta_i \theta_j}.
\ee
For conciseness we wrote it in terms of Grassmann integrals over $\theta_i$ and $\tilde{\theta}_i$, which can be expanded as a degree-$\lfloor \!\frac{n-2}{2}\! \rfloor$ polynomial in $\Lambda^2$ of Pfaffians, see, e.g., \cite[eq.~(4.8)]{Mizera:2019gea}. Similarly, we have the forms:
\be\label{varphi-bosonic}
\varphi^{\text{bosonic}}_{\pm} = d\mu_n\, (\pm\Lambda)^{n-2}\! \int \prod_{i=1}^{n} d\theta_i d\tilde\theta_i \exp \sum_{i \neq j} \Xi_{ij},
\ee
where
\be
\Xi_{ij} = \pm \frac{1}{\Lambda}\frac{ 2\theta_j\tilde\theta_j p_i {\cdot} \varepsilon_j}{z_i {-} z_j} + \frac{\theta_i \tilde\theta_i \theta_j \tilde\theta_j \varepsilon_i {\cdot} \varepsilon_j}{(z_i {-} z_j)^2}.
\ee
Upon the identification $\Lambda^2 {=} 1/\alpha'$, \eqref{varphi-gauge} and \eqref{varphi-bosonic} are in fact the same objects as those in super- and bosonic string perturbation theory respectively \cite{green1988superstring}, but---surprisingly---now appear in a purely field-theoretic context.

A partial list of theories whose amplitudes are known to have an interpretation as intersection numbers is given below \cite{Mizera:2019gea}:
\begin{center}
	\begin{tabular}{lll}
		$\varphi_-$ \qquad\qquad& $\varphi_+$ \qquad\qquad& theory\\
		\hline
		$\varphi_-^{\text{color}}$ & $\varphi_+^{\text{color}}$ & bi-adjoint scalar \cite{BjerrumBohr:2012mg,Cachazo:2013iea} \\
		$\varphi_-^{\text{color}}$ & $\varphi_+^{\text{gauge}}$ & Yang--Mills \\
		$\varphi_-^{\text{gauge}}$ & $\varphi_+^{\text{gauge}}$ & Einstein gravity \\
		$\varphi_-^{\text{color}}$ & $\varphi_+^{\text{bosonic}}$ & YM+$(DF)^2$  \cite{Johansson:2017srf,Azevedo:2018dgo} \\
		$\varphi_-^{\text{gauge}}$ & $\varphi_+^{\text{bosonic}}$ & Weyl--Einstein gravity \cite{Johansson:2017srf,Azevedo:2018dgo} \\
	\end{tabular}
\end{center}
Even though \eqref{varphi-gauge} depends on $\Lambda$, this dependence drops out from the resulting amplitudes in the first three cases (it is not true for the last two) \cite{Mizera:2019gea,Mafra:2011nv}. Since amplitudes are written as bilinears in this representation, KLT relations between the above theories become simply a consequence of linear algebra. The total differential $0 \sim\allowbreak (d{\pm} dW\wedge) \varphi_{\pm,n-1}^{\text{color}}$ implies the fundamental BCJ relation \cite{Bern:2008qj}, as an extension of the arguments in \cite{Cachazo:2012uq}. Twisted forms for states lying in the low-energy spectrum of string theory, such as those involving fermions or mixed Einstein--Yang--Mills interactions, can be readily written down using the techniques discussed in \cite{Mizera:2019gea}, but we will not pursue it here. Below we will extend the table with a few additional entries.

Scattering amplitudes in such a representation can be computed exactly using recursion relations \cite{Mizera:2019gea}, however the resulting numerators do not come in a Jacobi-satisfying way. Instead, the localization formula \eqref{intersection-number} is known to arise as the $\Lambda^0$ order in the \emph{low-energy} ($\Lambda \to \infty$) expansion of intersection numbers \cite{Mizera:2019gea},
\be\label{expansion}
\la \varphi_- | \varphi_+ \ra_{dW} = \sum_{\Gamma} \frac{\Res_{v_\Gamma}(\varphi_-) \Res_{v_\Gamma}(\varphi_+)}{\prod_{e\in \Gamma} p_e^2} + {\cal O}(\Lambda^{-2}),
\ee
when $\varphi_{\pm}$ are independent of $\Lambda$.\footnote{In the massless limit ($\Lambda \to 0$) intersection numbers have another localization formula on the so-called scattering equations, $dW{=}0$, which at the leading order $\Lambda^0$ gives the Cachazo--He--Yuan (CHY) \cite{Cachazo:2013hca,Cachazo:2013iea} formulation of massless amplitudes, see \cite{Mizera:2017rqa,Mizera:2019gea} for details. Since Yang--Mills and Einstein gravity amplitudes are independent of $\Lambda$ to begin with, this limit is exact. Subleading corrections ${\cal O}(\Lambda^{2p \geq 2})$ are given by higher residue pairings \cite{saito1983higher,Mizera:2019vvs}.} However, with the exception of $\eqref{varphi-color}$, twisted forms given above are polynomials in $\Lambda^2$, which leads to mixing of different orders in \eqref{expansion}. To consistently extract the leading order $\Lambda^0$ with \eqref{expansion} one needs to first remove the $\Lambda$-dependence from twisted forms by a repeated use of \eqref{cohomology-relation}. Given that Yang--Mills and Einstein gravity amplitudes are independent of $\Lambda$, once this is done the terms of order ${\cal O}(\Lambda^{-2})$ are not present and the numerators are exact.

\section{Examples}

We proceed with two illustrative examples. In order to contain expressions within the margins of this letter we focus on the case $n=4$, where amplitudes with color degrees of freedom take the form
\be\label{A-4}
{\cal A}_4 = \frac{n_s c_s}{s} + \frac{n_t c_t}{t} + \frac{n_u c_u}{u},
\ee
with $s = (p_1{+}p_2)^2$, $t=(p_2{+}p_3)^2$, $u=(p_1{+}p_{3})^2$ and a single triple. Fixing the punctures $(z_1,z_2,z_3)$ leaves us with a single coordinate $z_4$ on ${\cal M}_{0,4}$. Evaluating color numerators using \eqref{varphi-color} for $n=4$ amounts to computing the residues:
\begin{align}
c_s &= \Res_{z_4 = z_3} (\varphi_{-,4}^{\text{color}}) = f^{c_1 c_2 b} f^{b c_3 c_4},\nn \\
c_t &= \Res_{z_4 = z_1} (\varphi_{-,4}^{\text{color}}) = f^{c_2 c_3 b} f^{b c_1 c_4}, \\
c_u &= \Res_{z_4 = z_2} (\varphi_{-,4}^{\text{color}}) = f^{c_3 c_1 b} f^{b c_2 c_4}, \nn
\end{align}
with the convention $f^{abc} = \Tr(T^a [T^b, T^c])$.
In this case the residue theorem implies the usual Jacobi identity $c_s {+} c_t {+} c_u = 0$. We consider kinematic numerators next.

\subsection{Non-Linear Sigma Model}

Before obtaining numerators in Yang--Mills theory, let us consider a toy model of the color-kinematics duality between massless non-linear sigma model (NLSM) and special Galileon amplitudes \cite{Cachazo:2014xea}. To this end we use replacement rules of \cite{Cachazo:2014xea,Cheung:2017yef}, after which the twisted form $\varphi_{\pm}^{\text{gauge}}$ undergoes a vast simplification and becomes:
\be\label{varphi-scalar}
\varphi_{\pm}^{\text{scalar}} = -\frac{d\mu_n}{(z_k {-} z_\ell)^2} \det \mathbf{P}_{[k\ell]},
\ee
where the subscript $[k\ell]$ instructs one to remove columns and rows labeled by $k$ and $\ell$ prior to taking the determinant. Entries of the matrix $\mathbf{P}$ are given by
\be
\mathbf{P}_{ij} = \begin{dcases}
	\frac{2p_i {\cdot} p_j}{ z_i {-} z_j} \quad& \text{for } i \neq j,\\
	-\sum_{l \neq i} \frac{2p_i {\cdot} p_l}{ z_i {-} z_l} \quad& \text{for } i = j.
	\end{dcases}
\ee
Note that $\varphi_\pm^{\text{scalar}}$ is independent of $\Lambda$ and \eqref{expansion} has no ${\cal O}(\Lambda^{-2})$ corrections. We can add the following entries to the previous table:
\begin{center}
	\begin{tabular}{lll}
		$\varphi_-$ \qquad\qquad& $\varphi_+$ \qquad\qquad& theory\\
		\hline
		$\varphi_-^{\text{color}}$ & $\varphi_+^{\text{scalar}}$ & NLSM \\
		$\varphi_-^{\text{scalar}}$ & $\varphi_+^{\text{scalar}}$ & special Galileon \cite{Cachazo:2014xea} \\
		$\varphi_-^{\text{scalar}}$ & $\varphi_+^{\text{gauge}}$ & Born--Infeld \\
	\end{tabular}
\end{center}
For instance, choosing $(k,\ell)=(1,2)$ for $n=4$ we have
\begin{align}
\varphi_{+,4}^{\text{scalar}} &= \frac{z_{13} z_{32}}{z_{12}} \left(\frac{u}{z_{31}}{+}\frac{t}{z_{32}} {+} \frac{s}{z_{34}} \right)\left(\frac{t}{z_{41}}{+}\frac{u}{z_{42}} {+} \frac{s}{z_{43}}\right)dz_4
\nn\\
&\quad -s^2 \frac{z_{13} z_{23}}{z_{12} z_{34}^2} dz_4,\label{varphi-scalar-4}
\end{align}
where $z_{ij} := z_i {-} z_j$. Since \eqref{expansion} truncates at the leading order, we have the set of kinematic numerators for NLSM:
\begin{align}
n_s &= \Res_{z_4 = z_3} (\varphi_{+,4}^{\text{scalar}}) = s^2 + 2st,\nn \\
n_t &= \Res_{z_4 = z_1} (\varphi_{+,4}^{\text{scalar}}) = t^2, \\
n_u &= \Res_{z_4 = z_2} (\varphi_{+,4}^{\text{scalar}}) = -u^2, \nn
\end{align}
which satisfy $n_s {+} n_t {+} n_u = 0$. Note that numerators are not unique. For example, different initial choices of $(k,\ell)$ in \eqref{varphi-scalar} lead to distinct sets of numerators. Amplitudes of the special Galileon theory are obtained by replacing $c_\Gamma \to n_\Gamma$ in \eqref{A-4}.

\subsection{Gauge Theory}

Let us consider numerators in Yang--Mills theory. Choosing $(k,\ell)=(1,2)$ for $n=4$ the twisted form \eqref{varphi-gauge} becomes:
\be\label{varphi-gauge-4a}
\varphi_+^{\text{gauge}} = z_{13} z_{23} \left( \Pf \mathbf{\Psi}_{[12]} - 4 \Lambda^2 \frac{\varepsilon_1 {\cdot} \varepsilon_2\, \varepsilon_3 {\cdot} \varepsilon_4}{z_{12} z_{34}^2} \right) dz_4.
\ee
Here $\mathbf{\Psi}$ is the matrix known from the CHY formalism \cite{Cachazo:2013hca} in the conventions of \cite{Mizera:2019gea}. In order to fix the issue with $\Lambda$-non-homogeneity we use \eqref{cohomology-relation} with
\be
\xi = 4 \Lambda^2 \varepsilon_1 {\cdot} \varepsilon_2 \, \varepsilon_3 {\cdot} \varepsilon_4 \frac{z_{13} z_{24}}{z_{12} z_{34}},
\ee
obtained by integrating minus the final term in \eqref{varphi-gauge-4a}. Adding $(d{+}dW\wedge)\xi$ to \eqref{varphi-gauge-4a} gives us a form cohomologous to \eqref{varphi-gauge-4a}, but independent of $\Lambda$:
\begin{align}\label{varphi-gauge-4}
\varphi_{+,4}^{\text{gauge}} &= z_{13} z_{23}  \Pf \mathbf{\Psi}_{[12]}\, dz_4\\
&\quad + 4 \varepsilon_1 {\cdot} \varepsilon_2 \, \varepsilon_3 {\cdot} \varepsilon_4 \left(\frac{t}{z_{41}}{+}\frac{u}{z_{42}} {+} \frac{s}{z_{43}} \right) \frac{z_{13} z_{24}}{z_{12} z_{34}} dz_4.\nn
\end{align}
Therefore the leading order in \eqref{expansion} computes the full Yang--Mills amplitude. Using this representation we find:
\begin{align}
n_s &= \Res_{z_4 = z_3} (\varphi_{+,4}^{\text{gauge}})\nn\\
&= 8 \varepsilon_{1,\mu} \varepsilon_{2,\nu} \varepsilon_{3,\rho} \varepsilon_{4,\tau} [\,
p_1{\cdot} p_2 (\eta ^{\mu  \rho}\eta^{\nu  \tau } \!{-} \eta ^{\mu  \tau} \eta^{\nu  \rho })\\
&\;\quad{-}p_2{\cdot} p_3 \eta ^{\mu  \nu} \eta^{\rho  \tau }
+(p_1^{\rho } p_2^{\tau }
{-}p_2^{\rho } p_1^{\tau })\eta ^{\mu  \nu }
+p_1^{\nu } p_3^{\tau } \eta^{\mu  \rho }\nn\\
&\;\quad{-}p_1^{\nu } p_4^{\rho }\eta ^{\mu  \tau }
{-}p_2^{\mu } p_3^{\tau}\eta ^{\nu  \rho }
{+}p_2^{\mu} p_4^{\rho }\eta ^{\nu  \tau }
{+}(p_3^{\mu } p_4^{\nu } {-} p_4^{\mu } p_3^{\nu } )\eta ^{\rho  \tau }],\nn\\
n_t &= \Res_{z_4 = z_1} (\varphi_{+,4}^{\text{gauge}})\nn\\
&= 8 \varepsilon_{1,\mu} \varepsilon_{2,\nu} \varepsilon_{3,\rho} \varepsilon_{4,\tau} [\,
p_1{\cdot} p_2 \eta ^{\mu  \tau}\eta^{\nu  \rho }
+p_2{\cdot} p_3 \eta ^{\mu  \nu} \eta^{\rho  \tau } \\
&\quad+p_2^{\rho }p_1^{\tau }\eta ^{\mu  \nu }
-p_3^{\nu } p_1^{\tau }\eta ^{\mu  \rho }
+(p_1^{\nu } p_4^{\rho } 
{-}p_4^{\nu } p_1^{\rho } )\eta ^{\mu  \tau } \nn\\
&\quad+(p_2^{\mu } p_3^{\tau } 
{-}p_3^{\mu } p_2^{\tau })\eta ^{\nu  \rho }
-p_4^{\mu } p_2^{\rho } \eta ^{\nu  \tau }
+p_4^{\mu } p_3^{\nu }\eta ^{\rho  \tau } 
],\nn\\
n_u &= \Res_{z_4 = z_2} (\varphi_{+,4}^{\text{gauge}})\nn\\
&= 8 \varepsilon_{1,\mu} \varepsilon_{2,\nu} \varepsilon_{3,\rho} \varepsilon_{4,\tau} [\,
{-}p_1{\cdot} p_2 \eta ^{\mu  \rho} \eta^{\nu  \tau }
-p_1^{\rho } p_2^{\tau } \eta ^{\mu  \nu }\\
&\quad+(
p_3^{\nu } p_1^{\tau }
{-}p_1^{\nu } p_3^{\tau }) \eta ^{\mu  \rho}
+p_4^{\nu } p_1^{\rho } \eta ^{\mu  \tau } 
+p_3^{\mu } p_2^{\tau } \eta ^{\nu  \rho }\nn\\
&\quad+(
p_4^{\mu } p_2^{\rho }
{-}p_2^{\mu } p_4^{\rho })\eta ^{\nu  \tau }
-p_3^{\mu } p_4^{\nu } \eta ^{\rho  \tau }
].\nn
\end{align}
One can check that $n_s {+} n_t {+} n_u = 0$ and the resulting amplitude \eqref{A-4} is gauge invariant. Scattering amplitude of four gravitons is obtained by replacing $c_\Gamma \to \tilde{n}_\Gamma$ (with $\tilde\varepsilon_i$ instead of $\varepsilon_i$) followed by a symmetrization of polarization tensors, $\varepsilon^{\mu\nu}_i = \varepsilon_i^{(\mu} \tilde\varepsilon_i^{\nu)}$.

\section{Conclusion}

In this letter we introduced a representation of tree-level scattering amplitudes that manifests color-kinematics duality. The problem of finding theories with Jacobi-satisfying numerators translates to a classification of twisted forms, which motivates further extension of their available catalog.

The amplitudes computed with \eqref{varphi-gauge} have a remarkable property of being $\Lambda$-independent, as expected for massless theories, despite the fact $\varphi_{\pm}^{\text{gauge}}$ is not. 
On the other hand, it was previously shown that intersection numbers of logarithmic forms are independent of $\Lambda$ \cite{matsumoto1998,Mizera:2017rqa}. Thus, one might suspect that once $\varphi_{\pm}^{\text{gauge}}$ is brought into a $\Lambda$-independent form (perhaps using the algorithms of \cite{Mafra:2011kj,Mafra:2011nv,Ochirov:2013xba,Schlotterer:2016cxa,Fu:2017uzt,Du:2017kpo,He:2018pol,He:2019drm}) it would become logarithmic, as is the case for the examples \eqref{varphi-gauge-4} and \eqref{varphi-scalar-4}.\footnote{Although any twisted form can be written as a logarithmic form \cite{saito1980theory}, it is a non-trivial question whether such a form is independent of $\Lambda$ and has no kinematic poles. This is true in pure spinor superspace \cite{Mafra:2011kj,Mafra:2011nv}.} The answer has to be proportional to $\Pf \mathbf{\Psi}_{[k\ell]}$ plus corrections polynomial in $\partial W / \partial z_i$ since the latter ought to vanish after taking the $\Lambda \to 0$ limit which, by \eqref{cohomology-relation}, imposes scattering equations $dW{=}0$, cf. \cite{Fu:2017uzt,Du:2017kpo}. Finding a closed-form expression for all $n$ remains an open problem, which is of both theoretical and practical importance.

Generalization to higher-loop order consists of two separate steps. The first is writing down the analogue of \eqref{intersection-number} in terms of $(3g{+}n{-}3)$-fold residues on genus-$g$ moduli spaces, which necessarily satisfy the kinematic Jacobi identity by the same arguments as for $g{=}0$, thus proving that there is no topological obstruction to imposing \eqref{KJR} at any loop order. The second step is finding appropriate twisted forms generalizing \eqref{varphi-gauge} that give rise to loop integrands of gauge and gravity theories. The latter problem needs to be considered in the light of the fact that projectedness of supermoduli spaces (which was implicitly assumed in deriving \eqref{varphi-gauge}) breaks down at genus five \cite{Donagi:2013dua}.

\section{Acknowledgments}

\begin{acknowledgments}
The author thanks Ricardo Monteiro, Radu Roiban, and Edward Witten for many useful comments. He gratefully acknowledges the funding provided by Carl P. Feinberg.

The author would like to thank Hadleigh Frost and Lionel Mason for sharing their parallel work \cite{Frost:2019fjn} containing certain overlap with this letter.
\end{acknowledgments}

\bibliography{references}
	
\end{document}